\documentclass[pra,pre,10pt,showpacs,reprint,floatfix,nofootinbib]{revtex4-1}

\usepackage{graphicx,amsmath,amssymb}
\usepackage[usenames,dvipsnames]{color}
\usepackage[latin1]{inputenc}
\usepackage{ulem}

\newcommand{\avg}[1]{\langle #1 \rangle}
\newcommand{\ahum}[1]{``#1''}

\newcommand{\zs}{z_S}
\newcommand{\zu}{z_U}

\newcommand{\ru}{\rho_U}
\newcommand{\zc}{z_{\rm cr}}
\newcommand{\rhoc}{\rho_{\rm cr}}
\newcommand{\opd}{P_{L,i}(\ru|\zu,\zs)}

\newcommand{\beq}{\begin{equation}}
\newcommand{\eeq}{\end{equation}}
\newcommand{\bea}{\begin{eqnarray}}
\newcommand{\eea}{\end{eqnarray}}
\newcommand{\bvec}{\left( \begin{array}{c}}
\newcommand{\evec}{\end{array} \right)}

\newcommand{\eq}[1]{Eq.(\ref{#1})}
\newcommand{\fig}[1]{Fig.~\ref{#1}}
\newcommand{\olcite}[1]{Ref.~\onlinecite{#1}}

\begin{document}

\title{Domain formation in membranes with quenched protein obstacles: Lateral 
heterogeneity and the connection to universality classes}

\author{T.~Fischer and R.L.C.~Vink}

\affiliation{Institute of Theoretical Physics, Georg-August-Universit\"at
G\"ottingen, Friedrich-Hund-Platz~1, D-37077 G\"ottingen, Germany}

\begin{abstract} We show that lateral fluidity in membranes containing quenched 
protein obstacles belongs to the universality class of the two-dimensional 
random-field Ising model. The main feature of this class is the absence of a 
phase transition: there is no critical point, and macroscopic domain formation 
does not occur. Instead, there is only one phase. This phase is highly 
heterogeneous, with a structure consisting of micro-domains. The presence of 
quenched protein obstacles thus provides a mechanism to stabilize lipid rafts in 
equilibrium. Crucial for two-dimensional random-field Ising universality is that 
the obstacles are randomly distributed, and have a preferred affinity to one of 
the lipid species. When these conditions are not met, standard Ising or diluted 
Ising universality apply. In these cases, a critical point does exist, marking 
the onset toward macroscopic demixing. \end{abstract}



\maketitle

\section{Introduction}

Membranes are two-dimensional (2D) fluid environments \cite{citeulike:1595800}. 
There is growing consensus that membrane lateral structure is heterogeneous, and 
characterized by domains of different size and composition 
\cite{citeulike:5470515, citeulike:4308274, citeulike:4312704, 
citeulike:3042594, citeulike:6562069}. Domain formation in biological membranes 
is important because it links to key processes in cells, such as signaling, 
endocytosis, and adhesion \cite{citeulike:6505558, citeulike:6499105}, while in 
model membranes domain formation is relevant for applications, ranging from 
photolithographic patterning, spatial addressing, microcontact printing, and 
microfluidic patterning \cite{citeulike:6528926, citeulike:6549481}. To identify 
the factors that control domain formation, and to understand the underlying 
physical mechanisms, is therefore of practical importance.

In thermal equilibrium, a heterogeneous structure is difficult to comprehend, 
due to the large cost in line tension \cite{citeulike:5470515}. One instance 
where a heterogeneous structure does arise is near a critical point, as was 
demonstrated in free-floating giant unilamellar vesicles (GUVs) 
\cite{citeulike:3850776, citeulike:3850871, citeulike:3367620}. Near a critical 
point, the line tension vanishes; thermal motions then induce composition 
fluctuations over a wide range of length scales. An important concept in the 
theory of critical phenomena \cite{citeulike:1881924, citeulike:3367620} is 
universality: systems that belong to the same universality class undergo similar 
phase transitions, and yield the same set of critical exponents. For GUVs, the 
universality class was shown to be that of the 2D Ising model 
\cite{citeulike:3850776}. 

Since the universality class does not depend on the microscopic details of a 
system, it is tempting to speculate that 2D Ising universality, as observed in 
GUVs, is the generic class for membrane fluidity. The purpose of this paper is 
to show that, in less-idealized membranes (compared to GUVs), a different 
universality class comes into play, the reason being the presence of quenched 
obstacles. In living cells, proteins can bind to the underlying cytoskeleton 
\cite{citeulike:918597}. This leads to a 2D fluid consisting of mobile particles 
(e.g.~lipids) diffusing in a background of quenched (immobilized) protein 
obstacles. A similar situation arises in supported membranes, where surface 
friction may lead to particle immobilization \cite{citeulike:6127454}. Hence, to 
understand domain formation in membranes, it is important to keep in mind that 
an immobilized component may be present (in physical terms, such a system is 
called a quenched-annealed mixture \cite{madden.glandt:1988}). Precisely this 
point was recognized by two recent simulation studies \cite{citeulike:6599228, 
citeulike:7115548} where domain formation in membranes with quenched protein 
obstacles was investigated. It was found that quenched obstacles lower the 
critical temperature \cite{citeulike:6599228, citeulike:7115548}.

The aim of this paper is to relate the findings of 
Refs.~\onlinecite{citeulike:6599228, citeulike:7115548} to universality classes. 
Our main message is that, in the presence of quenched obstacles, membrane 
fluidity belongs to the universality class of the 2D random-field Ising model 
(2D-RFIM). The 2D-RFIM is crucially different from the 2D Ising model because it 
does not feature a phase transition \cite{citeulike:7028853, physrevb.23.287}. 
Consequently, critical behavior and macroscopic phase separation do not occur, 
even in the limit of low obstacle concentration. This means that, irrespective 
of temperature and lipid composition, a membrane with quenched obstacles is 
always in the same thermodynamic phase. The structure of this phase is found to 
be heterogeneous, consisting of micro-domains. Hence, based on universality 
alone, one elegantly accounts for a heterogeneous {\it equilibrium} domain 
structure over a wide range of compositions and temperatures. In contrast, 
critical fluctuations persist only in a small region around the critical point.

The universality class of the 2D-RFIM applies when the protein obstacles display 
a preferred affinity to one of the lipid phases, and are randomly distributed. 
We believe this to be the typical situation in most membranes. When these 
conditions are not met, standard 2D Ising (as observed in GUVs) or diluted 2D 
Ising universality \cite{citeulike:7927328} will arise. However, deviations 
between the latter two classes are relatively small, in the sense that both 
feature a critical point, with similar critical exponents 
\cite{citeulike:3506574, citeulike:4197235}.

\section{Model and Methods}

To illustrate these points, computer simulations of a 2D {\it off-lattice} 
mixture with quenched obstacles are performed. Since we focus on universality -- 
which does not depend on microscopic details -- it suffices to use a simple 
model. The protein obstacles are unit diameter disks, and are placed at the 
start of each simulation run on a $L \times L$ square with periodic boundaries, 
after which they remain quenched. The obstacles are mostly distributed randomly, 
although non-random choices will also be investigated. In line with recent 
simulations of membrane criticality \cite{citeulike:3850776, citeulike:3850871, 
citeulike:3850881, citeulike:3367620} a two-state model is used to describe the 
lipids, which diffuse through the environment of quenched obstacles. The model 
contains saturated (S) and unsaturated (U) lipids, both of which are disks 
having the same diameter as the obstacles. The sole interaction between lipids 
is a hard-core repulsion between unlike species, which is a minimum condition to 
induce phase separation \cite{widom.rowlinson:1970}. In addition, the lipids 
interact with the quenched obstacles; these interactions will be specified 
later.

In ternary mixtures of saturated and unsaturated lipids with cholesterol, 
critical behavior occurs over a wide range of compositions and temperatures 
\cite{citeulike:8017057}. To induce critical behavior, one either varies the 
temperature at fixed composition, or one varies the composition at fixed 
temperature; both routes are experimentally accessible \cite{citeulike:4145660}. 
Since the interaction in our model is hard-core, we vary the lipid composition. 
This is done in grand-canonical Monte Carlo (MC) simulations, i.e.~the 
respective fugacities $\zs$ and $\zu$ of saturated and unsaturated lipids are 
fixed, but the lipid number densities $\rho_X = N_X/L^2$ fluctuate, with $N_X$ 
the number of lipids of species $X \in (U,S)$ in the system (see 
\olcite{citeulike:7134501} for full details on the simulation method). The phase 
behavior is analyzed using the order parameter distribution (OPD) $\opd$, 
defined as the probability to observe a state with unsaturated lipid density 
$\ru$. The OPD depends on the imposed fugacities, the system size $L$, as well 
as on the configuration~$i$ of quenched obstacles. Since universality applies to 
the thermodynamic limit $L \to \infty$, meaningful results require simulation 
data over a range of~$L$, and the use of finite-size scaling (FSS) 
\cite{citeulike:1920630} to perform the extrapolation $L \to \infty$. In 
addition, in the presence of quenched obstacles, the shape of the OPD may 
fluctuate profoundly between obstacle configurations. It is therefore crucial to 
average simulation results over many $i=1, \ldots, K$ different obstacle 
configurations.

\section{Results}

\subsection{pure lipid membrane: 2D Ising universality}
\label{sec:NO}

\begin{figure}
\begin{center}
\includegraphics[width=0.8\columnwidth]{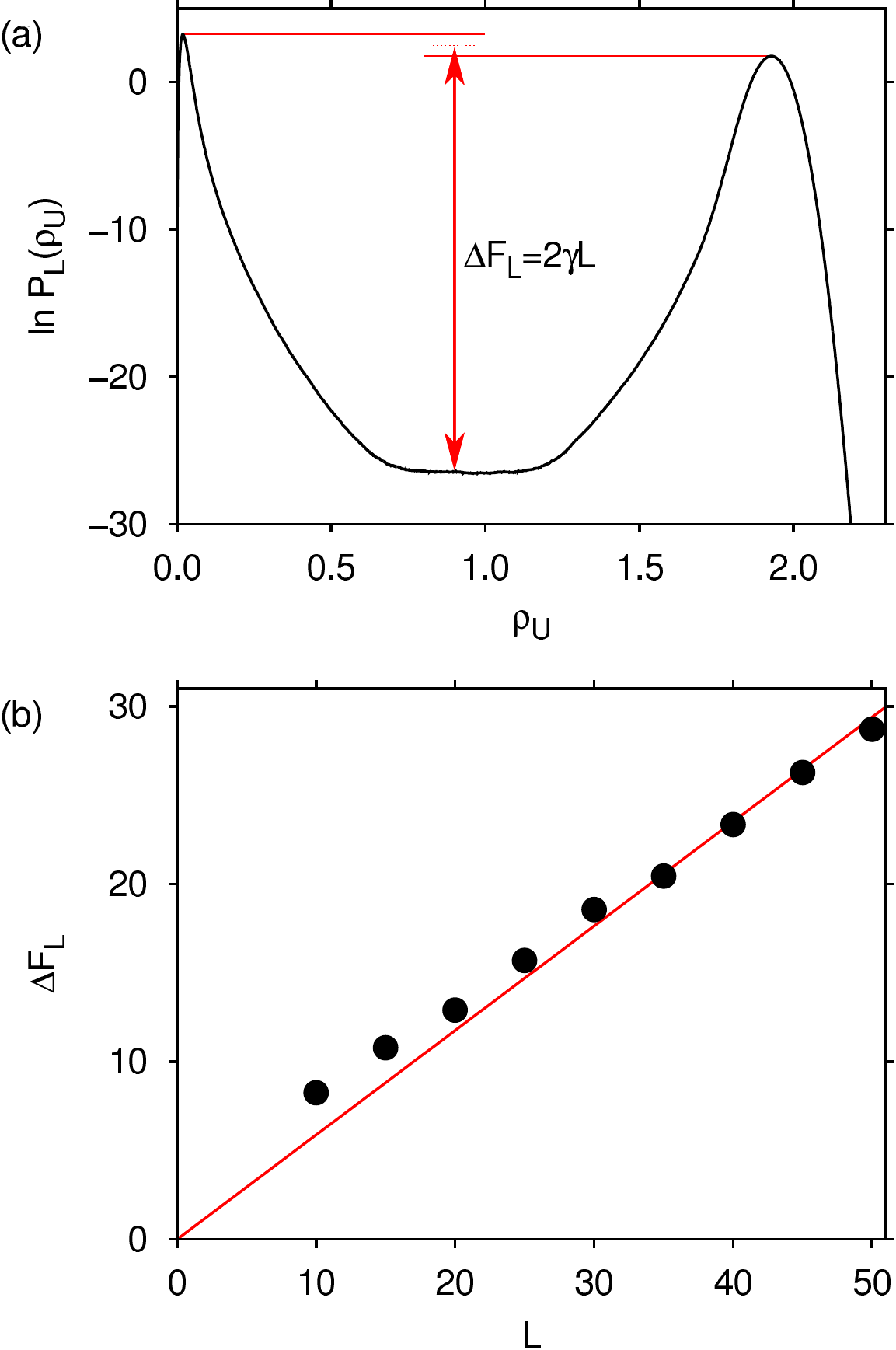} \\
\includegraphics[width=0.3\columnwidth]{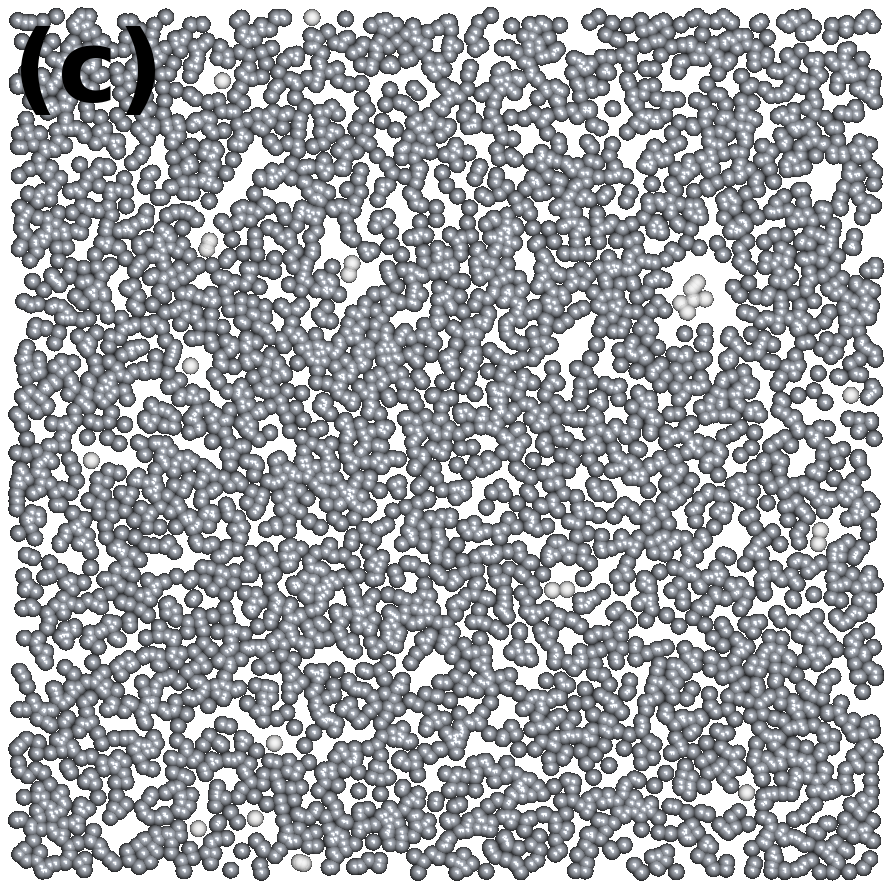}
\includegraphics[width=0.3\columnwidth]{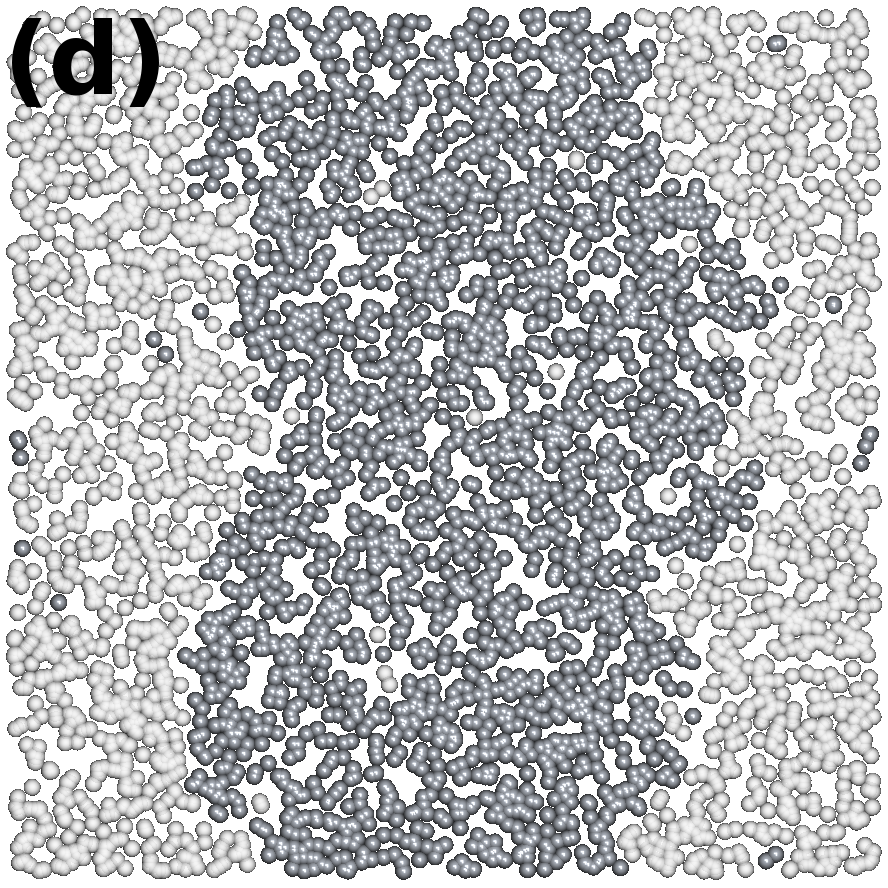}
\includegraphics[width=0.3\columnwidth]{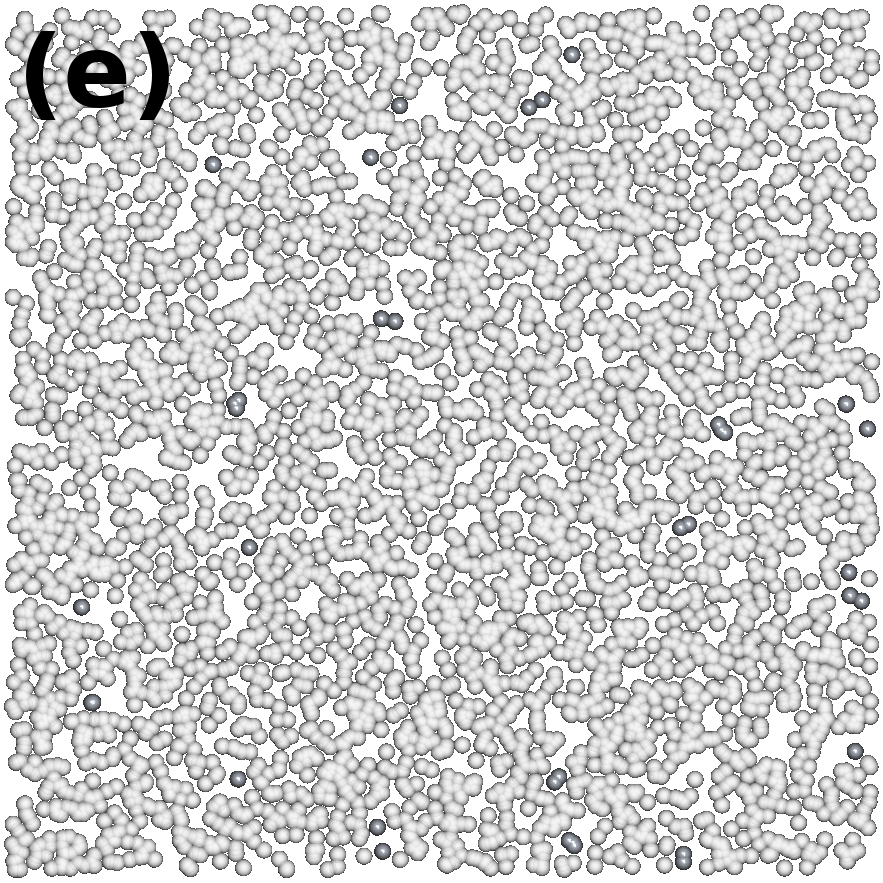}
\caption{\label{fig1} Finite-size effects in a pure lipid membrane at fugacity 
$\zs=2$ where the demixing transition is first-order. (a) The OPD for $L=50$, 
and with $\zu$ tuned according to \eq{eq:cx}. The distribution is distinctly 
bimodal. The vertical arrow marks the free energy barrier $\Delta F_L$ of 
interface formation. (b) The variation of $\Delta F_L$ with $L$. The linear 
increase confirms \eq{eq:lt} of a genuine first-order transition; from the slope 
of the line the line tension can be extracted. The lower frames show snapshots 
of the membrane corresponding to the left peak of the OPD (c), the coexistence 
region between the peaks (d), and the right peak~(e), where dark (light) 
particles represent saturated (unsaturated) lipids.}
\end{center}
\end{figure}

\begin{figure}
\begin{center}
\includegraphics[width=0.9\columnwidth]{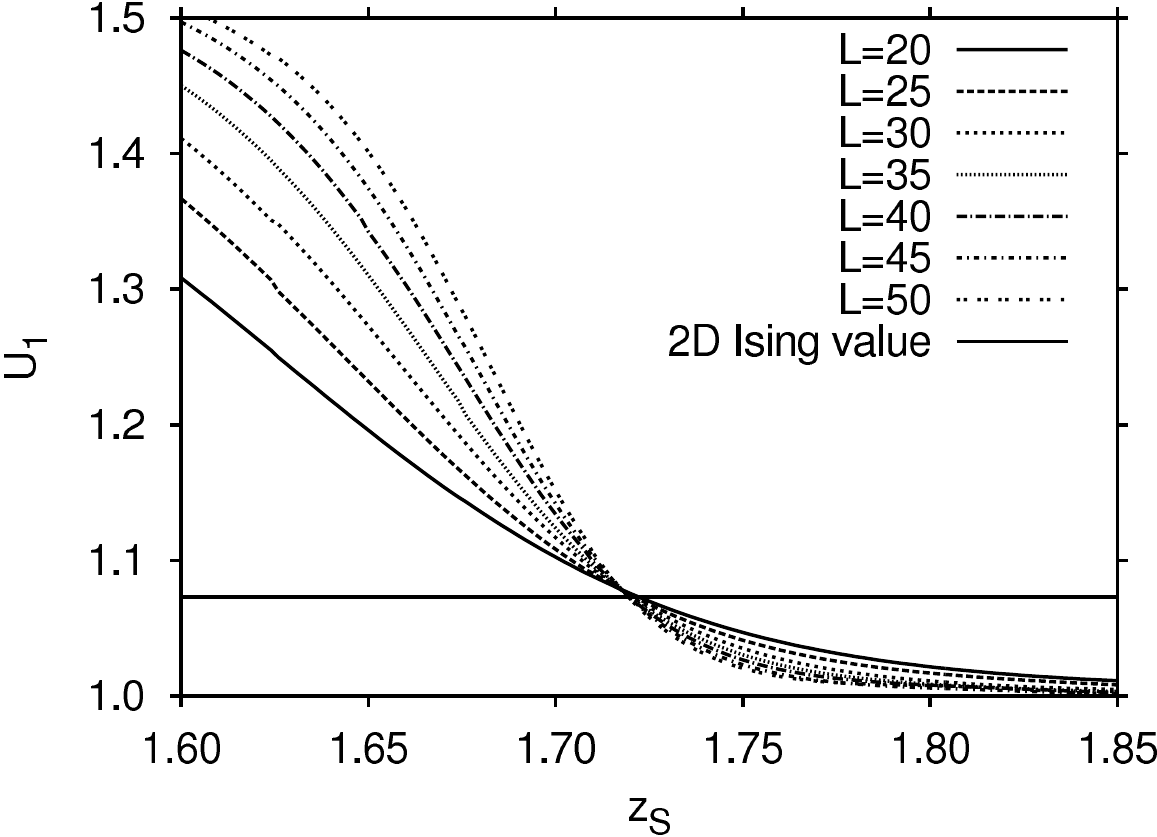}

\caption{\label{fig2} Demonstration of the cumulant intersection method 
\cite{binder:1981, citeulike:6170526} to locate the critical point in a membrane 
without quenched obstacles. Plotted is the cumulant $U_1$ versus the saturated 
lipid fugacity $\zs$ for several system sizes $L$, and $\zu$ tuned according to 
\eq{eq:cx}. At the critical point, the curves for different $L$ intersect. The 
horizontal line marks $U_1^\star \approx 1.07$ of the 2D Ising universality 
class.}

\end{center}
\end{figure}

We first consider a pure lipid membrane, i.e.~without quenched obstacles (this 
qualitatively resembles a GUV). We thus simulate saturated and unsaturated 
lipids only, both species being mobile. The control parameters are the particle 
fugacities, $\zs$ and $\zu$, and by appropriately tuning these parameters, phase 
transitions (provided they exist) can be induced. In some sense, the pure lipid 
membrane may be conceived as an {\it off-lattice} version of the 2D Ising model, 
with the role of temperature replaced by $1/\zs$, and that of the external field 
by $\ln \zu$. When $\zs$ exceeds a critical value $\zs>\zc$, the transition is 
first-order and macroscopic phase separation is observed. When $\zs=\zc$, the 
transition becomes continuous, and critical behavior is observed. When 
$\zs<\zc$, there is only one phase. At a phase transition, the scaling of the 
OPD with~$L$ assumes a characteristic form, from which the transition type can 
be determined. The observation of a transition also requires that $\zu$ is 
chosen suitably. In what follows, for a given value of $\zs$, we tune $\zu$ such 
that the derivative of the average saturated lipid density with respect to $\zu$ 
is maximized
\begin{equation}\label{eq:cx}
 \zu : \frac{ \partial \avg{\ru} }{ \partial \ln \zu } \to \mbox{max}, 
\end{equation} 
as is commonly done in fluid simulations \cite{orkoulas.fisher.ea:2001}. In the 
above, $\avg{\cdot}$ is a thermal average, i.e.~an integral over the OPD: $
\avg{\ru} = \int_0^\infty \ru \, P_L(\ru|\zu,\zs) \, d \ru$. 

At a first-order transition the OPD is bimodal; see \fig{fig1}(a), where the 
natural logarithm of the distribution is shown (there is obviously no dependence 
on the obstacle configuration~$i$ in this case). Provided the bimodal shape 
persists in the thermodynamic limit $L \to \infty$, the peaks reflect stable 
phases. The peak on the right corresponds to a homogeneous phase rich in 
unsaturated lipids, and lean in saturated lipids (\fig{fig1}(c)). The left peak 
corresponds to a homogeneous phase of reversed composition (\fig{fig1}(e)), 
i.e.~rich in saturated lipids, and thus resembles a lipid raft 
\cite{citeulike:2308151, citeulike:3042594, citeulike:6469836}. However, rafts 
are defined to be micro-domains, while the peaks in the OPD reflect macroscopic 
phases.

To verify that the bimodal shape survives in the thermodynamic limit, we 
consider the variation of the peak height $\Delta F_L$ with $L$ (vertical arrow 
in \fig{fig1}(a)). We emphasize that $\Delta F_L$ is obtained from the natural 
logarithm of the OPD: it is defined as the average peak height, measured from 
the minimum \ahum{in-between} the peaks. When the simulation traverses the 
region between the peaks, phase coexistence is observed. Both phases then appear 
simultaneously. Exactly between the peaks, each phase occupies half the system: 
the phases then arrange in two slabs since this yields the shortest interface 
(\fig{fig1}(d)). Note that, due to periodic boundaries, two interfaces are 
present. If $L$ is large enough, the interfaces do not interact with each other: 
the relative amount of the phases can then be varied over some range without any 
cost in free energy, which is the origin of the characteristic flat region 
between the peaks in the distribution of \fig{fig1}(a) \cite{citeulike:7237424}. 
Provided a flat region between the peaks is present, $\Delta F_L$ corresponds to 
the free energy cost of interface formation \cite{binder:1982}. Since, in the 
slab arrangement, the total interface length $l_{\rm slab} = 2L$, it follows 
that
 \begin{equation}\label{eq:lt}
  \Delta F_L = \gamma l_{\rm slab} = 2 \gamma L, 
 \end{equation}
with $\gamma$ the line tension \cite{binder:1982, physrevlett.65.137}. At a 
first-order transition in 2D, we thus expect a linear increase of $\Delta F_L$ 
with system size. This is confirmed in \fig{fig1}(b), and from the slope of the 
line $\gamma$ can be determined. Note that, for small $L$, corrections to 
\eq{eq:lt} become important \cite{billoire.neuhaus.ea:1994}, as indicated by the 
systematic deviation of the simulation data away from a straight line for $L<25$ 
or so.

Precisely at $\zs=\zc$, the first-order transition terminates in a critical 
point. The hallmark of criticality is scale invariance. This property can be 
exploited in FSS to extract $\zc$ from simulation data \cite{binder:1981, 
citeulike:6170526}. To this end, one measures the Binder cumulant $U_1 = 
\avg{m^2} / \avg{|m|}^2$, $m=\ru -\avg{\ru}$, which in the thermodynamic limit 
assumes three distinct values \cite{citeulike:1920630}. For $\zs>\zc$, 
i.e.~where the transition is first-order and characterized by two-phase 
coexistence, the OPD is bimodal (\fig{fig1}(a)). The OPD may then be 
approximated by a superposition of two non-overlapping Gaussian peaks 
\cite{citeulike:3717210}, which can be shown to yield $U_1=1$. For $\zs<\zc$, 
there is only one phase, with the OPD, consequently, featuring just a single 
peak. The peak is again Gaussian, for which $U_1=\pi/2$. Precisely at the 
critical point, the OPD is bimodal, but the peaks overlap. As a result, the 
Binder cumulant at criticality assumes an \ahum{in-between} value $1 < U_1^\star 
< \pi/2$; scale invariance implies that $U_1^\star$ does not depend on $L$. 
Moreover, $U_1^\star$ is characteristic of the universality class, and for the 
2D Ising model $U_1^\star \approx 1.07$.

In the thermodynamic limit, the Binder cumulant thus equals:
\begin{equation}\label{eq:u1}
 \lim_{L \to \infty} U_1 = \begin{cases}
 1 & \zs>\zc \hspace{5mm} \mbox{(two-phase region)}, \\
 U_1^\star & \zs=\zc \hspace{5mm} \mbox{(critical point)}, \\
 \pi/2 & \zs<\zc  \hspace{5mm} \mbox{(one-phase region)}.
 \end{cases}
\end{equation}
This behavior is well-suited to obtain the critical fugacity $\zc$ from 
simulation data. To this end, one plots $U_1$ versus $\zs$ for different system 
sizes (\fig{fig2}). At the critical point, the curves for different $L$ 
intersect. From the intersection point we conclude that $\zc \approx 1.72$ for 
the membrane model without quenched obstacles \cite{citeulike:7134501}. Note 
also that $U_1$ at the intersection is close to the 2D Ising value (horizontal 
line). Hence, from the Binder cumulant alone, we prove the existence of a 
critical point, obtain an estimate of the critical fugacity, and confirm 2D 
Ising universality. The critical fugacity, as well as the total lipid density at 
criticality $\rhoc \approx 1.56$ (which was extracted from the critical OPD), 
are in good agreement with other studies \cite{johnson.gould.ea:1997}. Of 
course, since the pure membrane model is symmetric under the inversion of $U 
\leftrightarrow S$ lipids, it trivially follows that $\rho_S = \rho_U$ at the 
critical point.

\subsection{random obstacles with preferred affinity: \\ 2D random-field Ising 
universality} \label{sec:RAO}

\begin{figure}
\begin{center}
\includegraphics[width=\columnwidth]{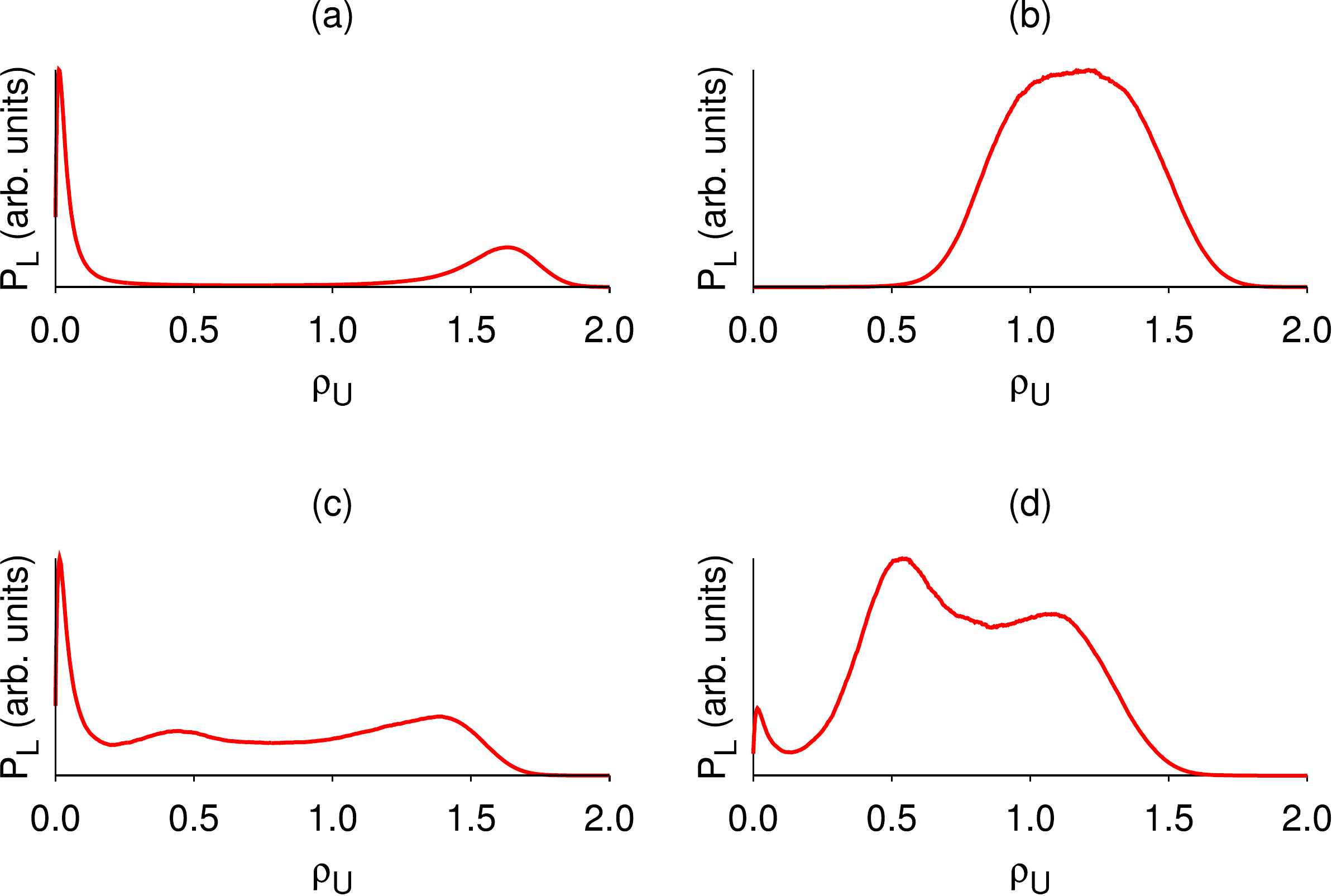}
\caption{\label{fig3} Some example OPDs for a membrane with quenched obstacles, 
with the obstacles randomly distributed, and with a preferred affinity to 
saturated lipids. The key observation is that the distributions profoundly 
fluctuate between obstacle configurations, which illustrates the need for an 
extensive average over many obstacle configurations; data are shown for $L=20$, 
$\zs=1.92$, and $\zu$ tuned according to \eq{eq:cx}.}
\end{center}
\end{figure}

\begin{figure}
\begin{center}
\includegraphics[width=\columnwidth]{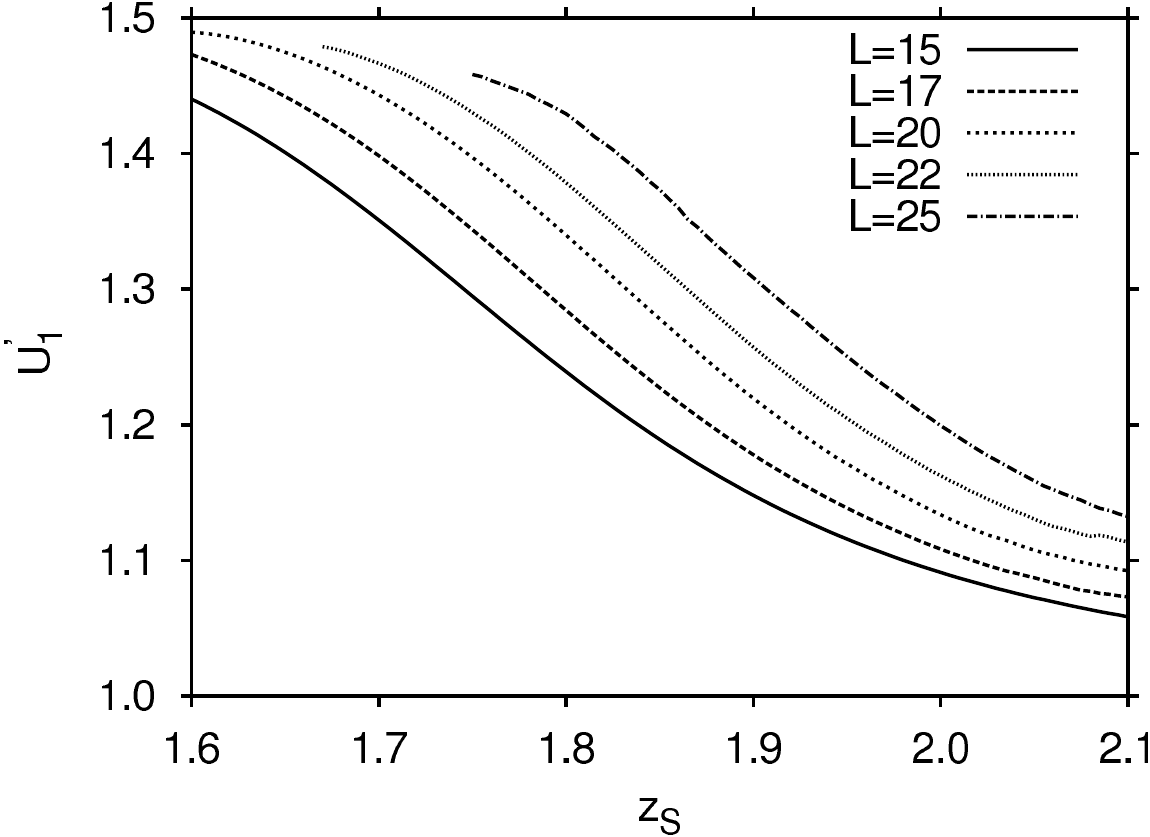}

\caption{\label{fig4} Cumulant analysis of a membrane with quenched obstacles, 
with the obstacles randomly distributed, and with a preferred affinity to 
saturated lipids. Plotted is $U_1'$ versus $z_S$ for different $L$. An 
intersection point is not observed: this indicates the absence of a phase 
transition, which is consistent with the universality class of the 2D-RFIM.}

\end{center}
\end{figure}

We now come to the main result of this paper, where we consider a membrane with 
quenched protein obstacles. The obstacles have a hard-core interaction with the 
unsaturated lipids, while saturated lipids may overlap freely with them. We thus 
impose a preferred affinity of the obstacles to lipid rafts, as, for example, 
GPI anchored proteins are known to do \cite{citeulike:2308151}. At the start of 
each simulation, we place $Q$ obstacles randomly in the system, after which they 
remain quenched; $Q$ is drawn from a Poisson distribution $P(Q) = (\lambda 
L^2)^Q e^{-\lambda L^2} / Q!$ with $\lambda = 0.03$ the average obstacle 
density. Next, saturated and unsaturated lipids are introduced, and the OPD is 
measured for the given configuration of quenched proteins. This procedure is 
repeated for many different obstacle configurations.

Since the proteins have a preferred affinity to one of the lipid species, and 
since they are randomly distributed, we expect 2D-RFIM universality. In 
particular: there should no longer be a phase transition. In \fig{fig3}, OPDs 
for a number of obstacle configurations are shown. The distributions were 
obtained for $\zs=1.92$, which significantly exceeds $\zc$ of the membrane 
without quenched obstacles. We observe large shape variations. In (a) we see a 
bimodal distribution, which might be taken as evidence of a first-order 
transition. However, in (b) only a single peak is revealed, which rather 
reflects no transition at all. Finally, (c) and (d) show distributions with 
three peaks, suggesting a triple-point. We emphasize that all distributions in 
\fig{fig3} were obtained at the same \ahum{inverse temperature} $\zs$ and system 
size: only the obstacle configurations are different. Since the distributions 
show extreme shape variations between obstacle configurations, it is clear that 
meaningful results require an average over many obstacle configurations. One 
might object that the shape variations in \fig{fig3} merely reflect a 
finite-size artifact, and that an average over obstacle configurations is not 
needed in larger systems. This, however, is a dangerous assumption because 
random-field systems are notoriously non-self-averaging 
\cite{aharony.harris:1996, citeulike:3201143}.

We thus use $K=2000$ obstacle configurations in what follows, and perform a FSS 
analysis to obtain insight in the phase behavior in the thermodynamic limit. To 
this end, we consider the quenched-averaged cumulant $U_1' = [\avg{m^2}] / 
[\avg{|m|}^2]$, with $m$ and $\avg{\cdot}$ defined as before, and where 
$[\cdot]$ is an average over obstacle configurations. For each obstacle 
configuration~$i$, the OPD is tuned according to \eq{eq:cx} 
\cite{citeulike:7690917}, which is then used to compute the thermal averages 
$\avg{\cdot}_i$; the latter are subsequently averaged over the obstacle 
configurations $[\avg{\cdot}] = (1/K) \sum_{i=1}^K \avg{\cdot}_i$. In 
\fig{fig4}, we plot $U_1'$ versus $\zs$ for different $L$. In contrast to 
\fig{fig2}, we cannot identify an intersection point. Instead, for a fixed value 
of $\zs$, the trend is that, by increasing $L$, the cumulant approaches the 
value of the one-phase region: $\lim_{L \to \infty} U_1' \to \pi/2$. Hence, 
irrespective of $\zs$, in the thermodynamic limit $L \to \infty$ there is only 
one phase. In agreement with 2D-RFIM universality, a phase transition (above 
which coexistence between two macroscopic phases would occur) no longer takes 
place.

The reason macroscopic phase separation in the 2D-RFIM is prevented has a clear 
physical origin: in the presence of quenched obstacles the cost of line tension 
becomes negligible. To see this, consider a raft domain of radius $R$. The cost 
of line tension scales $\propto R^{d-1}$, with $d$ the spatial dimension. 
However, the raft will also encompass a number of obstacles; the typical number 
of obstacles in the domain is $\propto R^d$ but with Poissonian fluctuations 
$\propto R^{d/2}$. Hence, it is favorable for the raft to seek out those regions 
in the membrane where there is an excess of obstacles. The cost of line tension 
is then compensated by the quenched obstacle excess, since $d=2$ for membrane 
fluidity. Note that this reasoning is just the analogue of the Imry-Ma argument 
for random-field magnets \cite{imry.ma:1975}.

\begin{figure}
\begin{center}
\includegraphics[width=0.48\columnwidth]{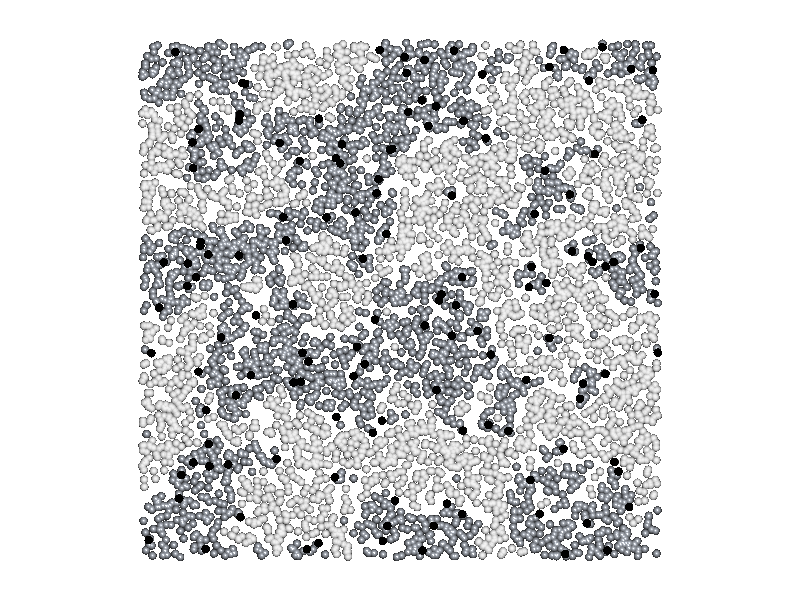}
\includegraphics[width=0.48\columnwidth]{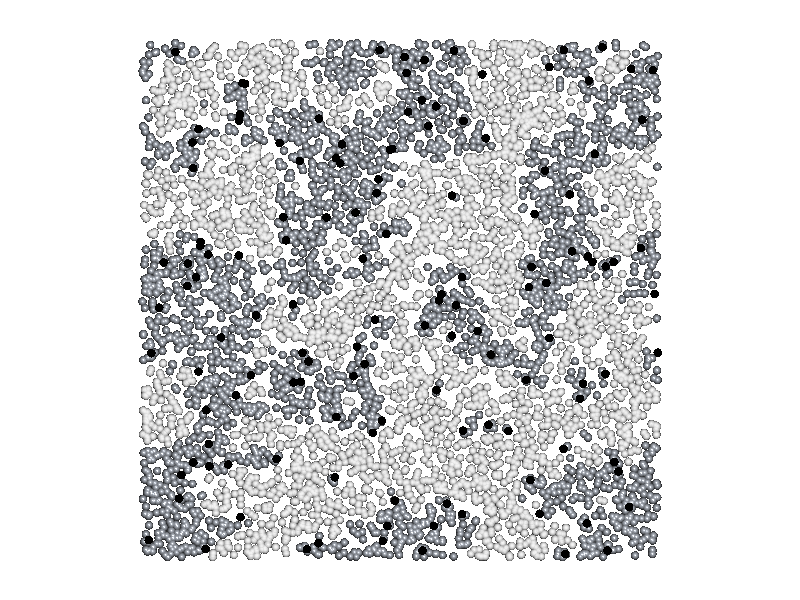} \\ 
\includegraphics[width=0.48\columnwidth]{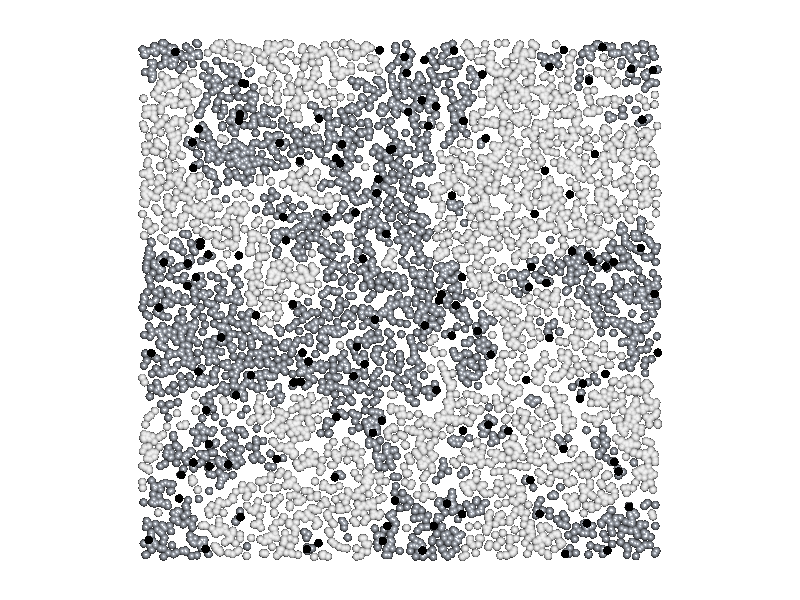}
\includegraphics[width=0.48\columnwidth]{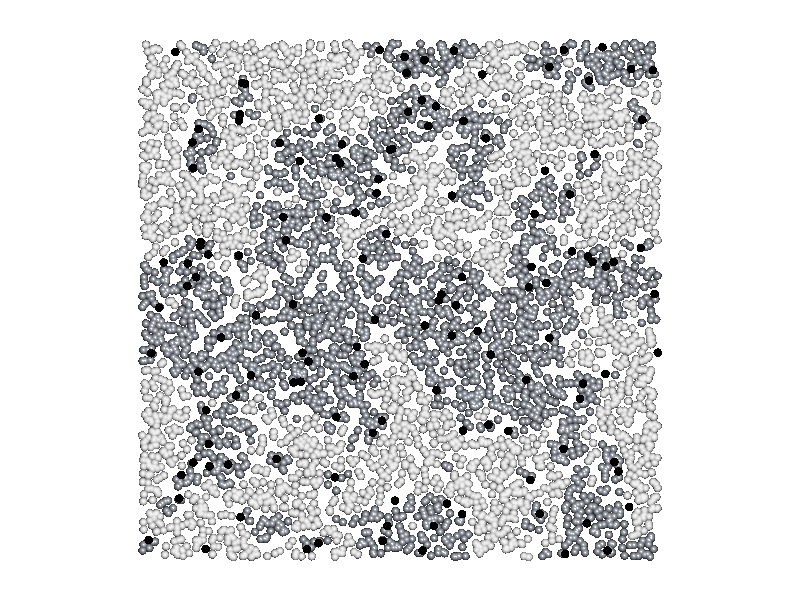} 
\caption{\label{fig:structure} Computer generated snapshots of domain formation 
in a membrane with quenched obstacles; $L=60$, $\rho=1.6$, and $\lambda=0.04$ 
were used. The dark (light) regions correspond to saturated (unsaturated) 
lipids; dots mark the protein obstacles.}
\end{center}
\end{figure}

\begin{figure*}
\begin{center}

\includegraphics[width=0.6\columnwidth]{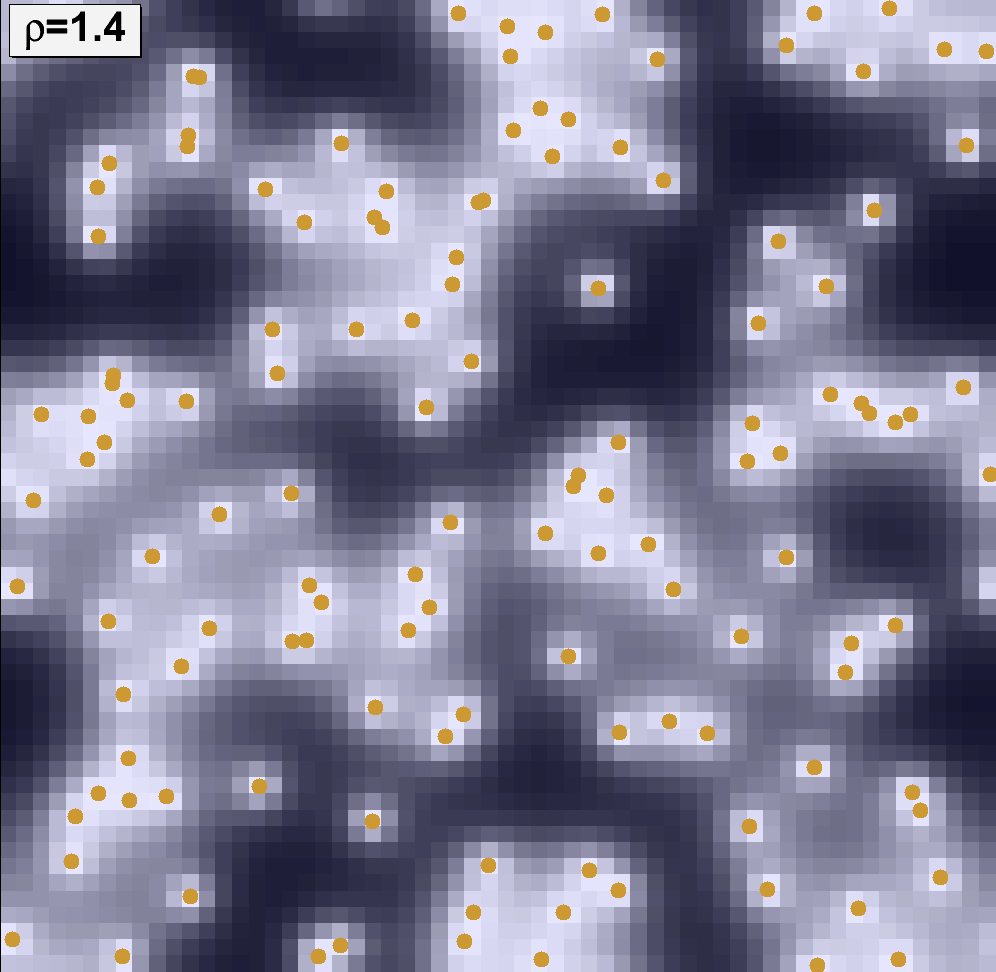} 
\includegraphics[width=0.6\columnwidth]{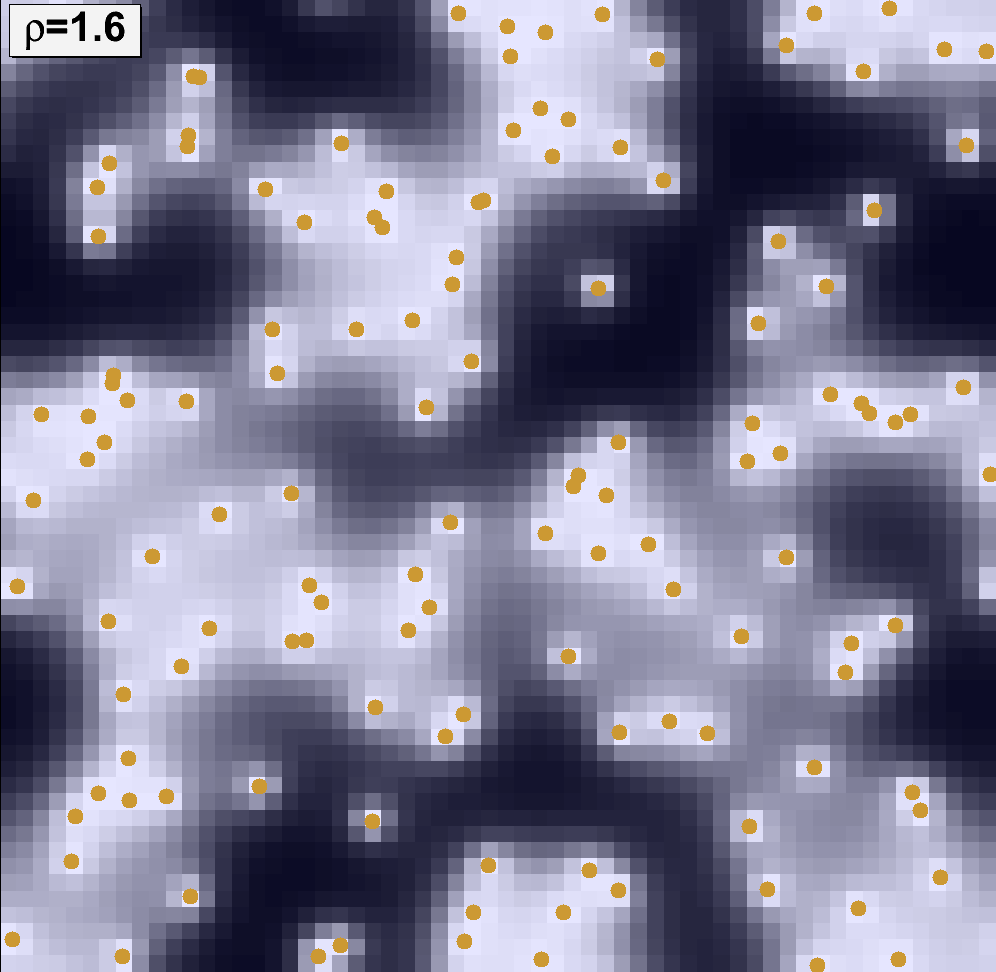} 
\includegraphics[width=0.6\columnwidth]{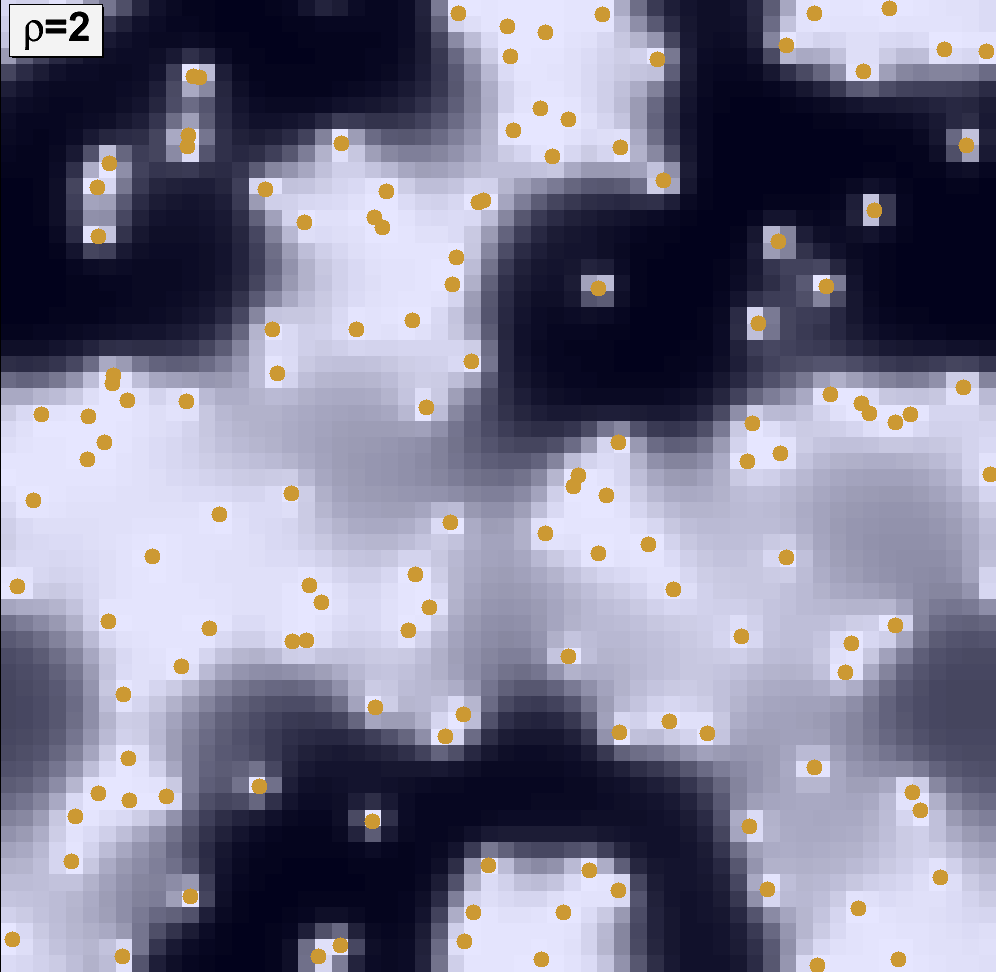}

\caption{\label{fig:Alocal} Series of {\it ensemble-averaged} snapshots of a 
membrane with quenched protein obstacles, for three values of the lipid density 
$\rho$, using $L=60$ and $\lambda=0.04$. The images are color-coded, and the 
color reflects the local lipid preference $A_k$ defined in \eq{def:A}, from 
$A_k=-1$ (brightest) to $A_k=1$ (darkest); dots mark the protein obstacles. Note 
that the same obstacle configuration is used in all three cases; for a different 
analysis where also an average over obstacle configurations is performed, see 
\fig{fig:Ahisto}. The images clearly show that the proteins create regions in 
the membrane where certain lipid species are preferred. In addition, the shape 
of these \ahum{preferred} regions is remarkably stable under variations in the 
lipid density.}

\end{center}
\end{figure*}

Having shown that randomly distributed quenched proteins (with preferred 
affinity to one of the lipid species) prevent macroscopic phase separation, the 
structure of the remaining single-phase will now be analyzed. To this end, we 
perform canonical (fixed density) MC simulations using $\rho_S = \rho_U$ for 
different total densities $\rho := \rho_S + \rho_U$. In \fig{fig:structure}, we 
show a number of equilibrated snapshots at $\rho=1.6$, $L=60$, and using the 
same obstacle configuration each time; the obstacle concentration 
$\lambda=0.04$. Note that $\rho=1.6$ exceeds the critical density of the pure 
membrane and so, if we were to remove the obstacles, the snapshots would reveal 
macroscopic phase separation, and resemble \fig{fig1}(d). However, from the 
cumulant analysis of \fig{fig3}, it is clear that this does not happen in the 
presence of quenched proteins: there is no longer a critical point, and hence no 
longer a two-phase coexistence region. Instead, we see a structure consisting of 
micro-domains (by micro we mean that the typical domain size exceeds the typical 
distance between the obstacles but remains finite). We also see that the domain 
structure in each of the snapshots is different: this shows that thermal 
fluctuations are still present, and that the domains have a finite lifetime 
(which is important because rafts are believed to be short-lived). At first 
sight, the domains in the snapshots of \fig{fig:structure} look deceivingly 
similar to critical fluctuations; see for example Fig.~1 of 
\olcite{citeulike:3850776} where critical fluctuations in GUVs are shown. 
However, the domains in \fig{fig:structure} are crucially different in two 
respects.

The first difference is that critical fluctuations are spatially indifferent: 
they can form at any location in the membrane with equal probability. In 
contrast, the domains that form in the presence of quenched obstacles are 
spatially selective: following the Imry-Ma argument, raft domains prefer to 
form at those locations in the membrane that feature an excess of obstacles. Of 
course, this spatial selection does not appear directly in single snapshots, but 
it becomes strikingly visible when we consider many snapshots and average over 
them. To demonstrate this explicitly, we use the same obstacle configuration of 
\fig{fig:structure}, and create a large number of equilibrated snapshots. For 
this set of snapshots, we construct an \ahum{ensemble-averaged} snapshot, simply 
by overlaying the individual snapshots. To this end, a grid consisting of unit 
square cells is placed over each individual snapshot. For each grid cell~$k$, we 
sum over all snapshots, and count how often the cell contained a saturated 
lipid, or an unsaturated lipid; the counts are denoted $C_S(k)$ and $C_U(k)$, 
respectively. From these counts, we compute the canonical (or time) averaged 
lipid preference for each cell
\beq \label{def:A}
 A_k := \frac{ C_U(k) - C_S(k) }{ C_U(k) + C_S(k) },
\eeq 
where a value $A_k=-1$ $(A_k=1)$ indicates a preference of cell~$k$ to saturated 
(unsaturated) lipids, while $A_k=0$ indicates that a preferred affinity is 
absent.

In case of critical fluctuations, which are spatially indifferent, the structure 
revealed in individual snapshots gets completely \ahum{washed-out} in the 
averaged snapshot, since $A_k=0$ for all cells in that case. In contrast, for 
the membrane with quenched obstacles, a clear structure in the averaged snapshot 
remains visible, which marks the regions in the membrane where rafts are most 
likely to be found. In \fig{fig:Alocal}, we show a number of averaged snapshots 
thus obtained, each one using the same obstacle configuration, but at different 
lipid densities $\rho$. The snapshots are color-coded, and the color reflects 
the value of $A_k$ in the grid cell; dots mark the quenched protein obstacles. 
For all lipid densities considered, which even includes one density slightly 
below the critical density of the pure model, it is clear that lipid domains are 
spatially selective.

The second difference is that critical fluctuations vanish above the critical 
density: the heterogeneous domain structure is then replaced by a coexistence 
between two (homogeneous) macro-domains. In contrast, the micro-domain structure 
in the membrane with quenched obstacles is stable upon variations in the total 
lipid density. The main effect of increasing $\rho$ is a \ahum{freezing-out} of 
thermal fluctuations. At high density, the membrane becomes more spatially 
selective, and it becomes less likely to observe raft domains in regions where 
there is no obstacle excess. This is manifested in the averaged snapshots of 
\fig{fig:Alocal} by a \ahum{sharpening} of the domain walls.

\begin{figure}
\begin{center}
\includegraphics[width=0.85\columnwidth]{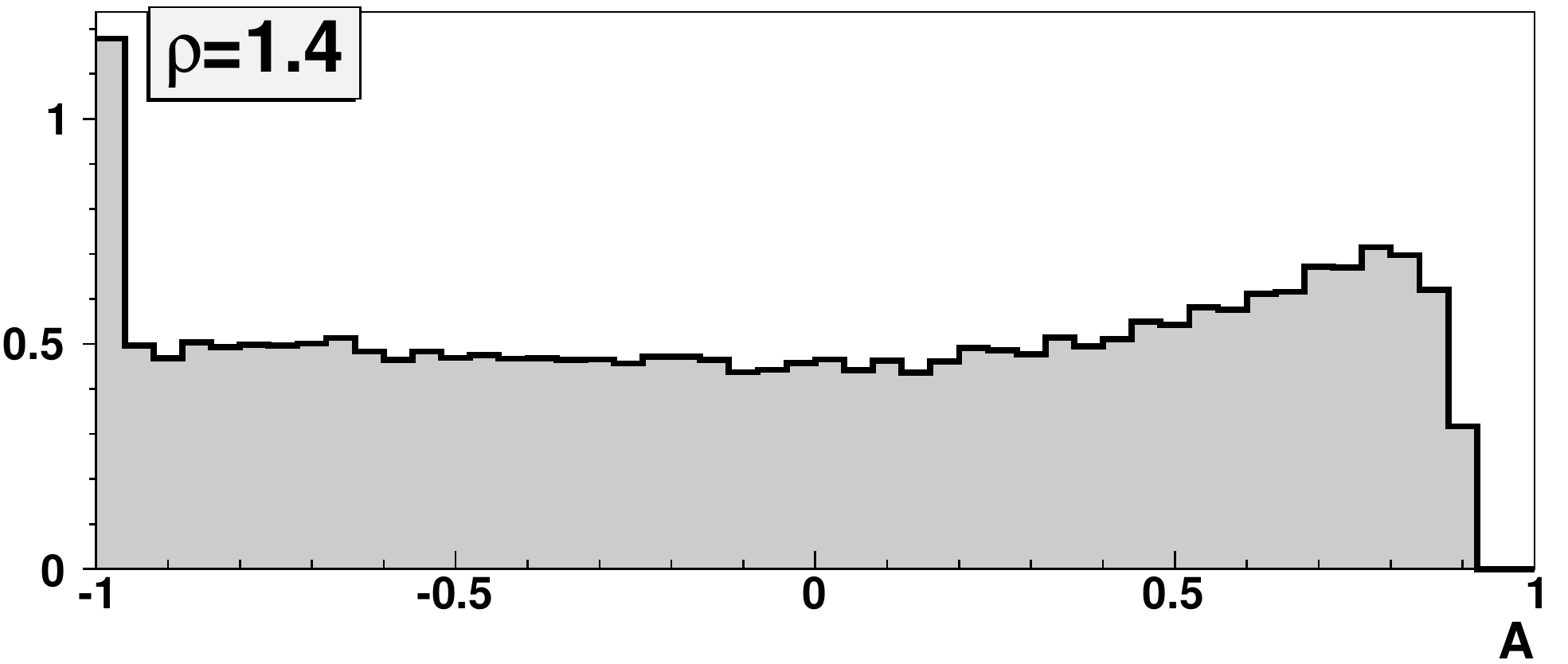} \\
\includegraphics[width=0.85\columnwidth]{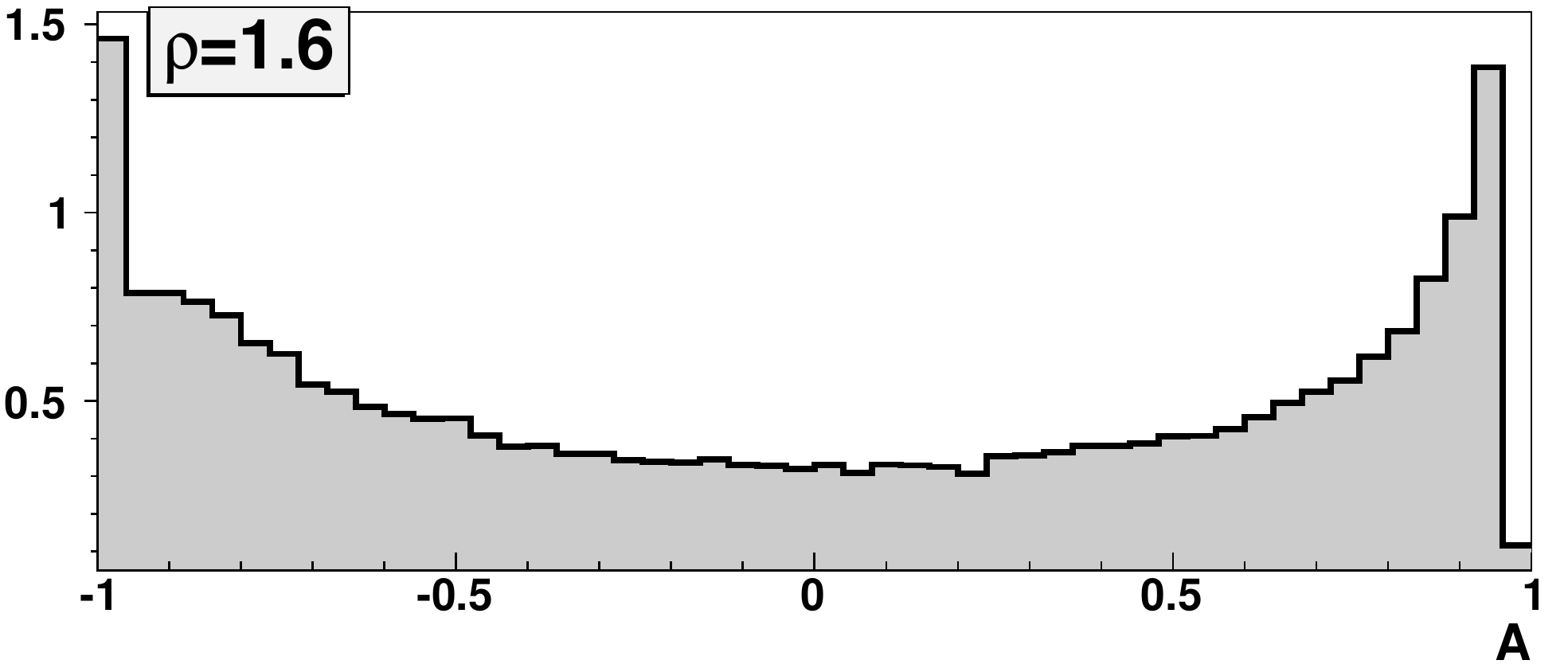} \\
\includegraphics[width=0.85\columnwidth]{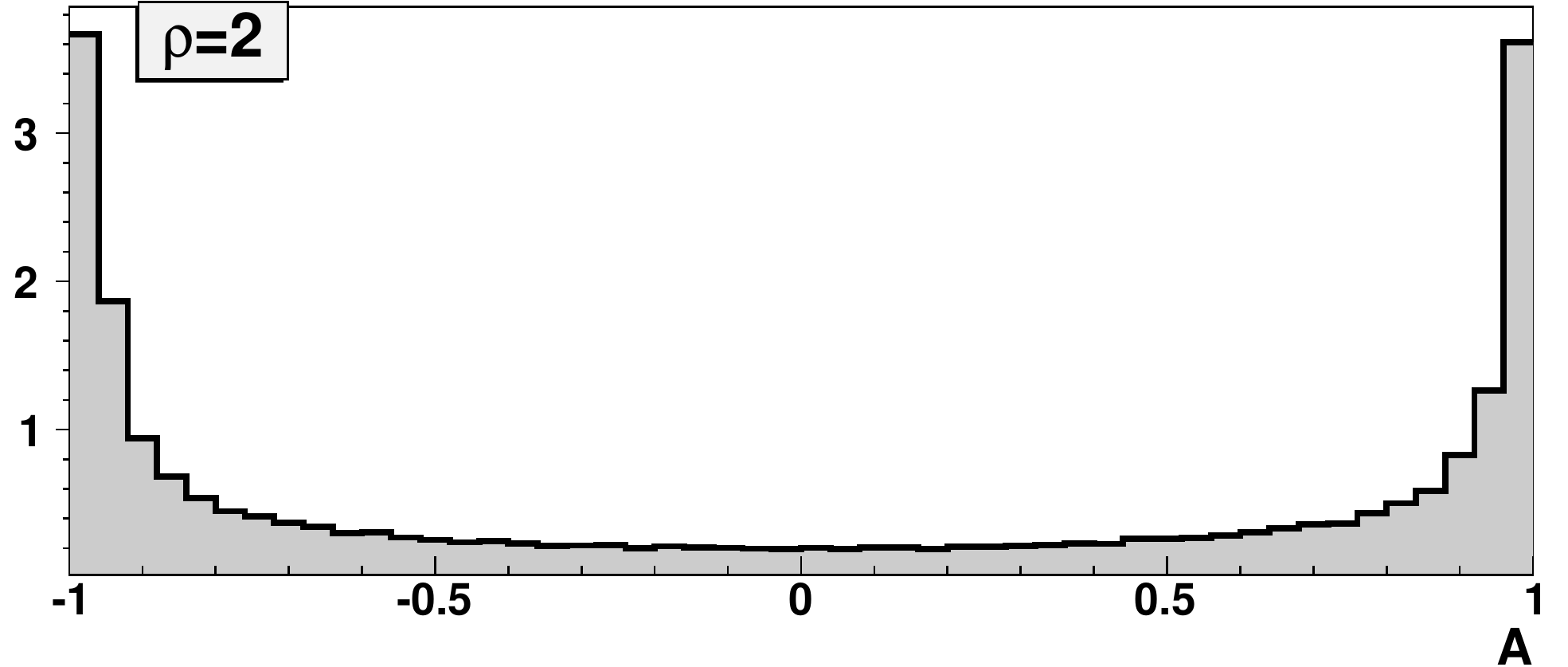} 
\end{center}

\caption{ \label{fig:Ahisto} Histograms of observed $A_k$ values for different 
values of the lipid density $\rho$ collected over 20 obstacle configurations.}

\end{figure}

This effect can be made more precise when we convert each averaged snapshot of 
\fig{fig:Alocal} into a one-dimensional histogram of $A_k$ values, and to 
subsequently average these histograms over a number of obstacle configurations. 
Of course, in this way we loose the spatial correlations, but it allows us to 
verify that the spatial selectivity shown in \fig{fig:Alocal} is a generic 
feature, and not just an artifact of the particular obstacle configuration that 
was used. In \fig{fig:Ahisto}, we show histograms of $A_k$ values thus obtained, 
which were averaged over~$20$ different protein configurations, and again for 
three lipid densities~$\rho$. Note that the sharp peak at $A_k=-1$, in 
particular for the $\rho=1.4$ histogram, is partially due to grid cells 
containing an obstacle. By increasing~$\rho$, the extreme values $A_k \sim \pm 
1$ become more likely, which confirms that the preference for certain lipids at 
certain locations in the membrane becomes more pronounced. The histograms also 
reveal another important point: even at density $\rho=1.4$, i.e.~below the 
critical density of the pure model, the membrane with quenched proteins is 
already spatially selective, since no peak around $A_k=0$ is visible.

{\it To summarize}: in the presence of quenched proteins, randomly distributed, 
and with a preferred affinity to one of the lipid species, there is no sign of a 
phase transition leading to two-phase coexistence, nor of a critical point. 
Instead of the lipids forming two macroscopic domains, micro-domains are 
observed. These micro-domains are dynamic, but they do not diffuse over the 
membrane randomly: they are most likely to be found in regions of the membrane 
featuring an excess of obstacles. In some sense, the quenched protein obstacles 
provide a scaffolding (channels) extending over the membrane for raft domains.

\subsection{random obstacles without preferred affinity : \\ diluted 2D Ising 
universality}\label{sec:RSO}

\begin{figure}
\begin{center}
\includegraphics[width=0.8\columnwidth]{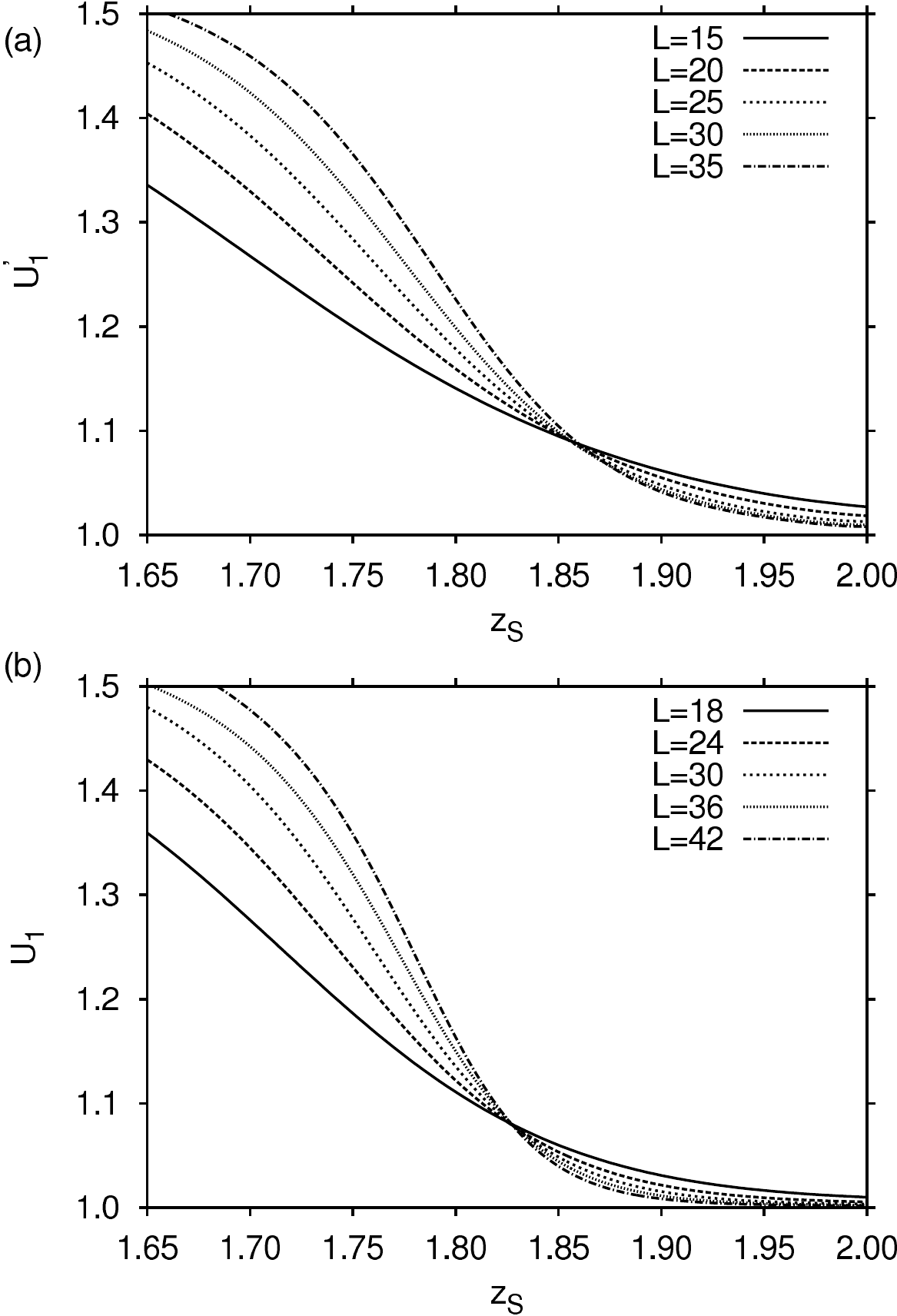} 
\caption{\label{fig5} (a) Cumulant analysis of a membrane with quenched protein 
obstacles that are randomly distributed, but without a preferred affinity to one 
of the lipid species. The curves of $U_1'$ versus $\zs$ for different $L$ reveal 
an intersection point, consistent with the occurrence of a phase transition. (b) 
The same analysis as above, but for protein obstacles with a preferred affinity 
to saturated lipids placed on a regular grid. Curves of $U_1$ versus $\zs$ for 
different $L$ intersect, consistent with the occurrence of a phase transition.}
\end{center}
\end{figure}

\begin{figure}
\begin{center}
\includegraphics[width=\columnwidth]{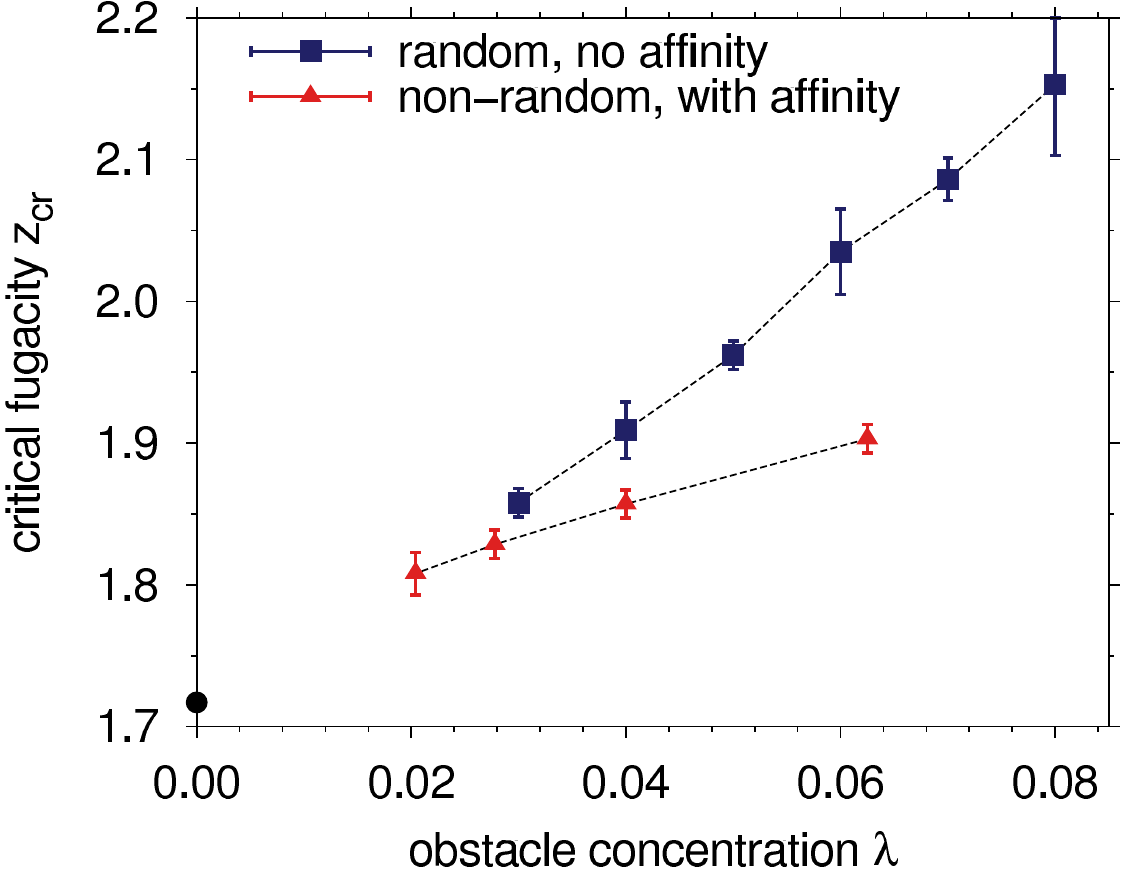}

\caption{\label{fig6} Variation of the critical fugacity $\zc$ with obstacle 
concentration $\lambda$ for a membrane with neutral obstacles that are randomly 
distributed (squares), and for a membrane with obstacles featuring a preferred 
affinity to saturated lipids, but placed on a regular grid (triangles). The dot 
at $\lambda=0$ marks $\zc$ of the pure membrane; lines serve to guide the eye.}

\end{center}
\end{figure}

We now consider a membrane with protein obstacles that do not have a preferred 
affinity to one of the lipid species, i.e.~are neutral, but remain randomly 
distributed. To this end, we use the rule that neither lipid species may overlap 
with the obstacles; the average obstacle concentration $\lambda=0.03$, again 
drawn from a Poisson distribution. The Imry-Ma argument does not apply in this 
case, and we indeed find a radical departure from 2D-RFIM universality. Instead, 
the scenario of \fig{fig1} is recovered, and a genuine phase transition is 
observed. In \fig{fig5}(a), we show the variation of the cumulant $U_1'$ with 
$\zs$, for different system sizes. An intersection point is revealed, proving 
the existence of a critical point, and for the critical fugacity we obtain $\zc 
\approx 1.86$. Since there is no longer a preferred affinity, the protein 
obstacles cannot compensate the cost of line tension. The membrane thus seeks to 
minimize the amount of interface, implying the formation of macroscopic domains 
when~$\zs>\zc$. Hence, in the presence of neutral obstacles, the only instance 
where a heterogeneous domain structure {\it in equilibrium} can arise is near 
the critical point. The corresponding critical fluctuations will be 
qualitatively similar to those of the pure membrane, i.e.~they are dynamic, and 
spatially indifferent (since the membrane with neutral obstacles remains 
symmetric under inversion of $U \leftrightarrow S$ lipids, it trivially follows 
that the averaged lipid preference $A_k=0$).

We emphasize that the case of neutral obstacles closely resembles the membrane 
simulations of \olcite{citeulike:6599228}. In that work, the 2D Ising model is 
studied, but with a fraction of randomly chosen lattice sites \ahum{turned-off} 
to represent quenched protein obstacles. The key point is that this kind of 
dilution disorder does not break the up/down symmetry of the Ising model, just 
as the neutral obstacles in our model do not break the $U \leftrightarrow S$ 
lipid symmetry. Consequently, both models belong to the same universality class, 
namely the one of the 2D diluted Ising model. For this universality class, there 
is no doubt that a phase transition exists \cite{citeulike:7927328}, provided 
the dilution remains below the limit where the lattice becomes disjoint. Indeed, 
an analysis of the Binder cumulant in \olcite{citeulike:6599228} also reveals an 
intersection point, in agreement with our \fig{fig5}(a). A second hallmark of 
diluted Ising universality is a pronounced decrease of the critical temperature 
with increasing obstacle concentration. This was strikingly confirmed in 
\olcite{citeulike:6599228}, and our data reveal the same trend (\fig{fig6}). 
Note that $\zs$ in our model plays the role of inverse temperature, so $\zc$ 
increases with the obstacle concentration. Finally, we point out that the 
extreme fluctuations of the OPD between obstacle configurations that 
characterize the random-field case (\fig{fig3}) do not occur in the diluted 
Ising model.

\subsection{non-random obstacles with preferred affinity : 2D Ising 
universality} \label{sec:GAO}

\begin{figure}
\begin{center}
\includegraphics[width=0.47\columnwidth]{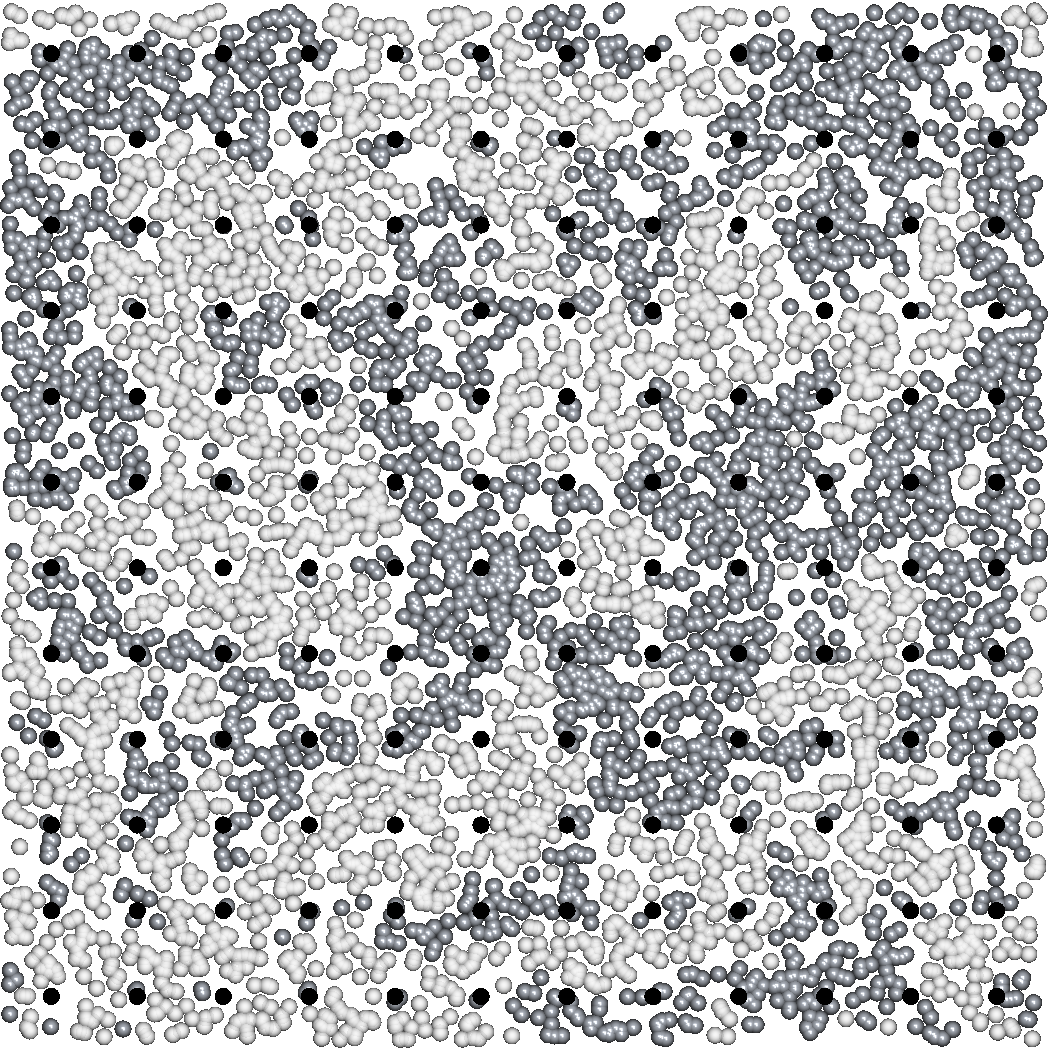}
\hspace{0.02\columnwidth}
\includegraphics[width=0.47\columnwidth]{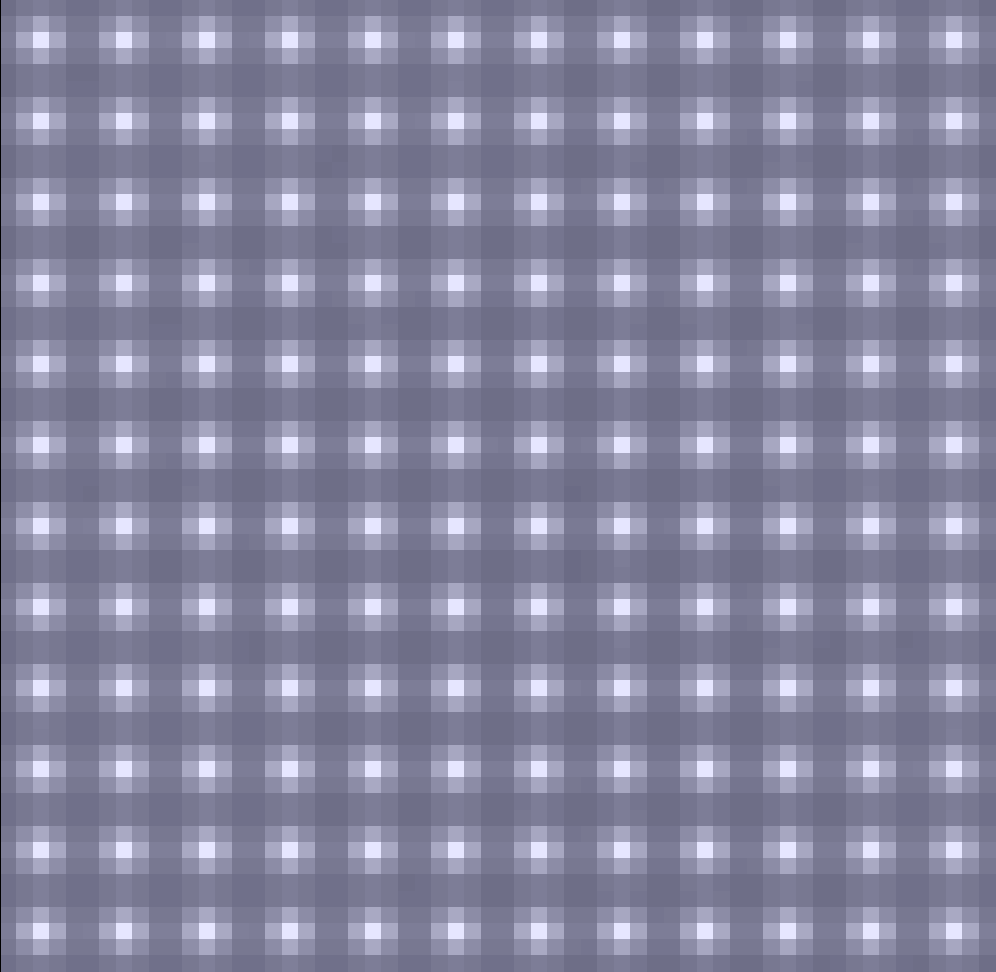}

\caption{\label{fig:GAO} Structure analysis of a membrane containing quenched 
protein obstacles; the obstacles have a preferred affinity to saturated lipids, 
and are placed on a grid with a lattice constant of five particle diameters. The 
left frame shows a single snapshot taken at densities slightly below the 
critical values $\rho_{U,\rm cr} \approx 0.82$ and $\rho_{S, \rm cr} \approx 
0.90$ (due to the preferred affinity, it does not hold that $\rho_{U,\rm cr} = 
\rho_{S, \rm cr}$). The right frame shows the corresponding {\it 
ensemble-averaged} snapshot. The main message is that the structure seen on the 
left is completely washed out. All that remains is a locally enhanced saturated 
lipid density around each obstacle.}
		
\end{center}
\end{figure}

Finally, we consider a membrane containing quenched proteins with a preferred 
affinity to saturated lipids, but placed non-randomly, namely on the sites of a 
square lattice. The lattice constant is six particle diameters, corresponding to 
an obstacle density $\lambda \approx 0.03$. In this case, the Imry-Ma argument 
does not apply either, because local (Poissonian) fluctuations in the obstacle 
density do not occur; note also that a disorder average $[\cdot]$ is not needed 
either. In \fig{fig5}(b), we show the cumulant $U_1$ versus $\zs$ for different 
$L$. We observe an intersection point, which proves the existence of a critical 
point, and for the critical fugacity $\zc \approx 1.83$ is obtained. Hence, the 
scenario of \fig{fig1} is recovered, with macroscopic phase separation taking 
place when $\zs>\zc$. By placing the protein obstacles at regular locations, the 
universality class remains that of the standard 2D Ising model. The only effect 
the proteins induce is confinement of the fluid, leading to a decrease of the 
critical temperature. Indeed, if we repeat the analysis using different lattice 
constants for the obstacle grid, we systematically find that $\zc$ increases 
with obstacle concentration (\fig{fig6}).

Hence, with the obstacles placed on a grid, an equilibrium heterogeneous domain 
structure only survives at the critical point. A typical snapshot obtained near 
criticality is given in the left frame of \fig{fig:GAO}, where now lattice 
constant five was used. Even though the obstacles have a preferred affinity to 
saturated lipids, this is not sufficient to stabilize a structure consisting of 
micro-domains. This becomes clear when one constructs the corresponding {\it 
ensemble-averaged} snapshot (conform \fig{fig:Alocal}) which we show in the 
right frame of \fig{fig:GAO}. In the averaged snapshot, the structure seen in 
the single snapshot is completely washed out. Of course, due to the preferred 
affinity to saturated lipids, there is a layer of enhanced saturated lipid 
density around each obstacle, but this layer is restricted to a single obstacle. 
These layers do not extend to other obstacles to form micro-domains (one might 
regard them as nano-domains instead). In contrast, the micro-domains in 
\fig{fig:Alocal} extend over numerous protein obstacles.

\subsection{Conclusion}

Inspired by recent simulations of membranes with quenched protein obstacles 
\cite{citeulike:6599228, citeulike:7115548}, additional simulations were 
performed with the aim to relate domain formation in these systems to 
universality classes. The main message is that, while a membrane without 
quenched obstacles is in the universality class of the 2D Ising model 
\cite{citeulike:3850776}, the introduction of obstacles will in most practical 
situations induce a change to 2D random-field Ising universality. The 
consequences are drastic: the phase transition of the pure membrane without 
obstacles is destroyed, macroscopic phase separation no longer occurs, and only 
a single phase survives consisting of micro-domains. These micro-domains prefer 
to form at special regions in the membrane, and this preference becomes more 
pronounced with increasing lipid density. It would be of great interest to see 
whether transport of particles happens along the same regions; in that case, 
quenched proteins truly provide a structure of channels along which membrane 
constituents are laterally transported.
	
We have also considered two somewhat \ahum{artificial} cases, whereby the 
quenched obstacles are neutral, or placed regularly on a grid. In these cases, a 
normal critical phase transition is observed, above which macroscopic phase 
separation takes place. These predictions are presumably less relevant for 
biological cells, but they could be verified in experiments on model membranes 
(using, for example, an appropriately patterned surface 
\cite{citeulike:6609138}). We write {\it presumably} above because the spectrin 
cytoskeleton network in red blood cells does, in fact, form a regular lattice 
structure \cite{citeulike:6116283}.

On a more fundamental level, the connection between a fluid with quenched 
obstacles (with preferred affinity) and random-fields dates back to de~Gennes 
\cite{gennes:1984}, but has been notoriously difficult to verify experimentally 
(see \olcite{citeulike:5845715} for a review). An alternative for the hypothesis 
of de~Gennes, which in some cases better captures experiments, is the 
\ahum{wetting hypothesis} of \olcite{citeulike:4878283}. In the latter 
description, randomness is not the deciding factor in the resulting phase 
behavior. However, it is clear that our simulations are only compatible with the 
random-field hypothesis of de~Gennes, since we see large differences between 
randomly and non-randomly placed obstacles.

Finally, having shown that domain formation in membranes with quenched obstacles 
belongs to the universality class of the 2D-RFIM, further insights might be 
gained by comparing to other systems in that class. The obvious candidates are 
filled polymer blends confined to thin films \cite{citeulike:6590578}, for which 
the connection to the random-field Ising model was already noted 
\cite{citeulike:6562394}. Experiments have revealed that quenched filler 
particles dramatically limit phase separation dynamics \cite{citeulike:6590482}. 
Complete phase separation is typically not observed, but rather a mosaic of 
microscopic domains \cite{citeulike:1615711}. Computer simulations and theory 
\cite{citeulike:6562394, citeulike:4149114, citeulike:4146944} indicate that the 
dynamics can become completely arrested, leading to many \ahum{pinned} domains 
\cite{citeulike:4219764, citeulike:6196444}. Some of these results are in 
remarkable agreement with the membrane simulations of 
Refs.~\onlinecite{citeulike:6599228, citeulike:7115548} and the present work.

\acknowledgments

This work was supported by the {\it Deutsche Forschungsgemeinschaft} (Emmy 
Noether program:~VI~483/1-1). We also acknowledge stimulating discussions at the 
CECAM workshop {\it Complex dynamics of fluids in disordered and crowded 
environments}, July 2010, where the results of this paper were first presented.

\end{document}